\documentclass[lettersize,journal]{IEEEtran}
\usepackage{amssymb,amsmath}
\usepackage{cite}
\usepackage{graphicx}
\usepackage{psfrag}
\usepackage{url}
\usepackage[latin1]{inputenc}
\usepackage[absolute,overlay]{textpos}
\usepackage{gensymb}
\usepackage{cases}
\usepackage{graphicx}
\usepackage[font=footnotesize]{caption}
\usepackage[font=footnotesize]{subcaption}
\usepackage{float}
\usepackage[linesnumbered,ruled,lined]{algorithm2e}
\usepackage{pgf}

\usepackage[nocomma]{optidef}

\usepackage{amsthm}

\usepackage{booktabs}
\usepackage{color,soul}

\begin{document}

\title{Multi-UAV Path Learning for Age and Power Optimization in IoT with UAV Battery Recharge}
\author{
	\IEEEauthorblockN{Eslam Eldeeb, Jean Michel de Souza Sant'Ana, Dian Echevarr\'ia P\'erez, Mohammad Shehab,\\ Nurul Huda Mahmood, and Hirley Alves\\
	}
	\thanks{This work has been partially supported by Academy of Finland 6G Flagship program (Grant no. 346208), FIREMAN (Grant no. 326301), and the European Commission through the Horizon Europe project Hexa-X (Grant Agreement no. 101015956).}
	
	\thanks{The authors are with Centre for Wireless Communications (CWC), University of Oulu, Finland. Email: firstname.lastname@oulu.fi.}
}
\maketitle

 \vspace{-2mm}
\begin{abstract}
In many emerging Internet of Things (IoT) applications, the freshness of the is an important design criterion. \textit{Age of Information} (AoI) quantifies the freshness of the received information or status update. This work considers a setup of deployed IoT devices in an IoT network; multiple unmanned aerial vehicles (UAVs) serve as mobile relay nodes between the sensors and the base station. We formulate an optimization problem to jointly plan the UAVs' trajectory, while minimizing the AoI of the received messages and the devices' energy consumption. The solution accounts for the UAVs' battery lifetime and flight time to recharging depots to ensure the UAVs' \textit{green operation}. The complex optimization problem is efficiently solved using a deep reinforcement learning algorithm. In particular, we propose a deep Q-network, which works as a function approximation to estimate the state-action value function. The proposed scheme is quick to converge and results in a lower ergodic age and ergodic energy consumption when compared with benchmark algorithms such as greedy algorithm (GA), nearest neighbour (NN), and random-walk (RW).
\end{abstract}
 \vspace{-1mm}
\section{Introduction}

The Internet of Things (IoT) era is allowing the implementation of new time-sensitive applications through the deployment of sensor nodes to collect information in real-time. Use cases include intelligent transportation, environmental monitoring, and human safety. To address time sensitivity in such applications, a metric termed as Age of Information (AoI) was introduced in~\cite{kaul2011minimizing} to quantify the degree of freshness of the information about a certain process. It is defined as the time elapsed since the generation of the packet that was most recently delivered to the destination node. 
The application of unmanned aerial vehicles (UAVs) as mobile relay units has been proved to be very efficient in solving the problem of minimizing the AoI while maintaining energy limitations~\cite{tang2020minimizing}. The UAV relays can reduce the transmission distance of IoT nodes by moving close to the source nodes and then relaying the transmitted information to the destination node \cite{mozaffari2019tutorial}. This facilitates communication and saves energy in remote areas, where it is cumbersome to replace the batteries of the sensor nodes.


Recently, learning schemes such as deep reinforcement learning (DRL) have been extensively applied in solving the problem of jointly minimizing the AoI and energy consumption in IoT. However, the suitability of a DRL algorithm is strongly conditioned on the dimension of action and state spaces, which turns out to be a curse in massive scenarios~\cite{DQNs}. This issue can be handled by deploying multiple UAVs to collect information along with device clustering to reduce the state-action spaces.


Several works have considered the use of UAV for AoI minimization. For instance, the authors in~\cite{abd2019deep} jointly optimized the scheduling policy and flight trajectory of the UAV to minimize the weighted sum AoI. The work in~\cite{9195789} proposed a DRL model to minimize the freshness of information in a single-hop vehicular network. In~\cite{9750860}, the authors presented a multi-agent DRL solution to coordinate between the UAVs to efficiently perform wireless energy transfer (WET) and wireless information transfer (WIT). To minimize the AoI in massive deployment up to fifty devices, the work in~\cite{samir2020online} presented a model-free DRL solution, whereas the authors in~\cite{ferdowsi2021neural} formulated the problem as a mixed-integer program and a convex-optimization-based solution.


To this end, the contributions of this paper are summarized as follows:
	\begin{itemize}
		\item We propose a DRL solution to jointly minimize the AoI and the devices energy consumption in a massive deployment of up to hundred IoT devices.
		\item Our model accounts for UAVs battery constraints and flying time to recharging depots.
		\item We apply k-means to perform device clustering, while accounting for the UAVs scheduling capacity.
		\item Our approach outperforms the baseline RW, greedy and NN models in terms of age and IoT energy consumption.
	\end{itemize}
 \vspace{-1mm}

\section{System Layout and Problem Formulation}\label{system}
\subsection{System Model}


We consider a 2D grid world of a set $\mathcal{K}=\{1,2,\cdots,K\}$ of $K$ low-power IoT devices. Each device is randomly distributed in the grid world and is given a coordinate $c_k=(x_k,y_k)$ after being projected to the 2D plane as in \cite{deep_us,deep_china}. The IoT devices are served by a set $\mathcal{U} =\{1,2,\cdots,U\}$ of $U$ rotary-wing UAVs. Each UAV flies over the grid world to collect information from the devices and relay the collected information to the BS located at the center of the grid world (i.e, at $(0,0)$). The grid world has fixed charging depots $D$ located at the four corners.

\begin{figure}[t!]
	\centering
	\includegraphics[width=0.95\columnwidth]{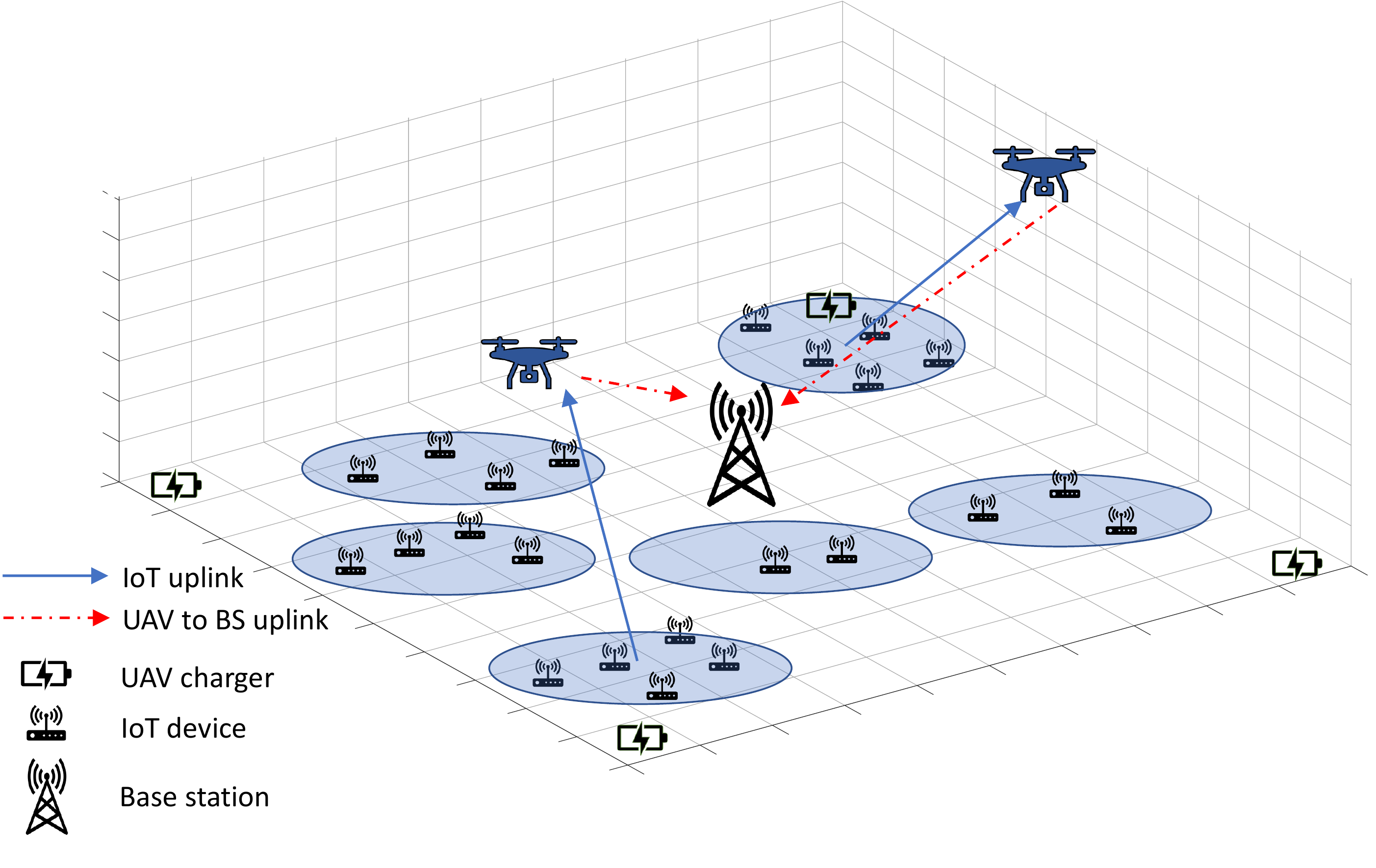}
	\caption{System model: IoT clusters are served by multiple UAVs. Each UAV relays the information from the IoT clusters to the BS in the middle of the map.} \vspace{-2mm}
	\label{Fig1}
\end{figure}

Each UAV starts and ends its trajectory at one of the charginf depots. The grid world is divided into square cells, where the movement of each UAV occurs in four directions (i.e, east, west, north, south) or preserving its location by not moving at all (hovering). Time slots are discretely divided as [$\tau$, $2\: \tau$, ...], where $\tau$ is the time that the UAV needs to move from the center of one cell to the center of an adjacent cell. The time unit $\tau$ is determined by calculating the ratio between the distance between the centers of two adjacent cells $d_g$ and the velocity of the UAV $\upsilon_t$. The system model is illustrated in Fig. \ref{Fig1}.  \vspace{-1mm}

 \vspace{-1mm}
\subsection{Preliminaries}
\subsubsection{Energy Consumption}\label{ENERGY_CALC}
Consider that the scheduling policy of the IoT devices $S(t) \in \mathcal{S} = \{0,1,...,K\}$, where $S(t) = (k_1, k_2, ...)$ means that the nodes $k_1, k_2, ...$ are scheduled to transmit at time slot $t$. Each UAV forwards the received packet to the BS. We assume the presence of LOS communication between the sensors and UAVs, and between the UAVs and BS, therefore, the channel gain between UAV $u$ and the BS at time slot $t$ is given by \vspace{-1mm}
\begin{align}
\label{eq_3}
    g_{u,BS}(t) = g_0d_{u,BS}^{-2} = \frac{g_0}{|h_u-h_{BS}|^2+||c_u(t)||^2},
\end{align}
where $g_0$ is the channel gain at the reference distance of 1 m, $d_{u,BS}$ is the distance between the UAV and the BS, $h_u$ is the altitude of the UAV, $h_{BS}$ represents the height of the antennas at the BS, and $c_u(t)$ is the position of UAV $u$ at time instant $t$\cite{deep_us}.
$P_k$ is the transmission power of an IoT device $k$ and it is calculated as follows
\begin{align}
    P_k = 
    \frac{(2^{\frac{M}{B}}-1)\sigma^2}{g_0}\:\Bigg(d_{u,k}^2 + h_u^2\Bigg),
\end{align}
where $M$ is the packet size of the sensor updates, $B$ defines the signal bandwidth, $\sigma^2$ the noise power, and $d_{u,k}$ is the distance between UAV $u$ and IoT device $k$ \cite{deep_china}.

We discretize the battery capacity of each UAV $E_{max,u}$ into energy quanta $\mathcal{N}_u$, where the amount of energy in each energy quantum is given by the ratio $E_{max,u}/\mathcal{N}_{u}$. Denote the battery level of UAV $u$ at time slot $t$ as $e_u(t) \in \mathcal{E}_u = \{0,1,...,e_{u,max}\}$. The battery of the UAV is affected by the energy consumed to relay an update packet to the BS $e_u^R(t)$ and the energy consumed due to flying or hovering $e_u^F(\upsilon_t)$. 
The battery evolution of the UAVs can be described as
\begin{equation}
	e_u{(t\!+\!1)}=
	\begin{cases}
		e_u(t)-\lceil e_u^R(t)+e_u^F(\upsilon_t) \rceil, & \text{if}   \ \mathcal{S}(t) = k, \\
		e_u(t)-\lceil e_u^F(\upsilon_t)\rceil, & \text{otherwise}, \end{cases}
\end{equation}
where $\lceil \: \rceil$ is ceiling approximation. The energy consumed to relay an update packet to the BS is given by
\begin{align}
    e_u^R(t) = \frac{\mathcal{N}_{u}}{E_{max,u}}E_u(t),
\end{align}  \vspace{-2mm}
with  \vspace{-2mm}
\begin{align}
    E_u(t) = \frac{\sigma^2}{g_{u,BS(t)}}\big(2^{\frac{M}{B}}-1\big),
\end{align}
whereas the energy consumed due to flying or hovering is given by  \vspace{-1mm}
\begin{equation}
    e_u^F(\upsilon_t)=\frac{\mathcal{N}_{u}}{E_{max,u}} P_u(\upsilon_t),
\end{equation}
where $P_u(\upsilon_t)$ is the power consumption of the UAVs when moving or hovering and is formulated in \cite{zeng2019energy} as
\begin{align}
P_u(\upsilon_t)=P_0& \left( 1+\frac{3\upsilon_t^2}{s_{tip}^2} \right)+P_1\left(\sqrt{1+\frac{\upsilon_t^4}{4s_0^4}}-\frac{\upsilon_t^2}{2s_0^2}\right)^\frac{1}{2} \nonumber\\
+&\frac{1}{2}d_0\rho \mu_0 Z \upsilon_t^3,
\end{align}
where $P_0$ and $P_1$ represent the blade profile power and derived power when the UAVs are hovering, respectively, $\upsilon_t$ describes the velocity of the UAVs and $S_{tip}$ depicts the tip speed of the blade. Meanwhile, $s_0$ is the mean rotor induced velocity when hovering, $d_0$ represents the fuselage drag radio, $\rho$ is the air density, $\mu_0$ represents the rotor solidity and $Z$ the area of the rotor disk.

\subsubsection{AoI Calculation}
We formulate the discrete AoI as the time elapsed since the last time a device transmitted a packet. The AoI is used as a degree of fairness in scheduling the devices. If a device transmits an update packet, its AoI is reset to one. The AoI of device $k$ is given by
\begin{equation}
\label{AOI_CALC}
	A_k(t+1) =
	\begin{cases}
		1, & \quad \text{if} \ \mathcal{S}(t) = k, \\
		\text{min}\{A_{max},A_{k}(t) + 1\}, & \quad \text{otherwise}, 
	\end{cases}
\end{equation}
where $A_{max}$ denotes the maximum allowed AoI in the model.
 \vspace{-3mm}

\subsection{Problem Formulation}\label{INF}
The main objective of the UAVs is to jointly minimize the weighted average AoI and the transmission power of the IoT devices. Hence, We the optimization problem is formulated as follows   \vspace{-1mm}
\begin{subequations}\label{P1}
	\begin{alignat}{2}
	\mathbf{P1:}\qquad &\underset{\boldsymbol{l}(t)}{\min}       &\ \ \ & \frac{1}{T}\sum_{t=1}^T\sum_{k = 1}^{K}\delta_k A_k(t) + \frac{\lambda}{K} \sum_{k = 1}^{K} P_k(t),\label{P1:a}
	\ \\
	&\text{s.t.}   &      & \sum_t^{T_u}P_u(\upsilon_t)\leq e_u(t), \label{P1:b}\\
		& & & c_u(1) = c_{d,u}, \label{P1:c}
	\end{alignat}
\end{subequations}
where $\delta_k$ is the importance weight that denotes the importance of device $k$ and $c_{d,u}$ are the coordinates of the charging depot where UAV $u$ is going to take off. Here, $\lambda$ is a multiplicative variable that controls the trade-off between the AoI and the transmission power. The larger the value of $\lambda$ the more the objective function cares about the power over the AoI. If $\lambda = 0$, the model learns to produce the best AoI without taking the transmission power into account. The constraints of the given optimization problem assure that the UAVs still have enough energy to move and serve the devices and forcing the initial and final positions of each UAV to be at one of the charging depots.

The optimization problem~\eqref{P1} is a non-linear integer programming optimization problem whose complexity grows with the number of deployed devices. In addition, the UAV experiences a large dimension of state space, which is almost a continuous state space. To overcome the dimensionality curse, we propose a DRL with a deep Q-network (DQN) approach, which works as a function approximation to estimate the Q-function and solve the given problem efficiently and feasibly.

\vspace{-2mm}
\section{The proposed DRL solution}\label{DQN}

\subsection{Clustering and Rate-Mobility Characterization}
Consider that each device k is assigned to a cluster $l \in \mathcal{L}$, where $\mathcal{L} = \{1,2,\dots, L\}$ is a set of clusters of length $n$. We call $n_l$ the number of devices on cluster $l$. A UAV will try to communicate with all devices within a cluster $l$ based on a given policy. For this, before starting moving from one grid position to another, the UAV will send an uplink grant to all devices in the specified cluster. Thus, devices should be able to transmit their updates before the UAV arrives at the next position. Hence, the relation between the number of devices on a cluster $n_l$ and the fixed transmission rate $R_{b_l}$ of devices cluster $l$  is given as $n_l \leq \frac{R_{b_l} \tau}{M}$. Thus, substituting $\tau=\frac{d_g}{\upsilon_t}$, we have  \vspace{-1mm}
\begin{equation}
    n_l \leq \frac{R_{b_l} d_g }{M \upsilon_t }.\label{eqn:rate_1}
\end{equation}


Note that this number is directly related to the average rate, and the speed of the UAV. The BS performs the clustering using k-means according to the positions of the devices and by setting the calculated maximum number of devices in a cluster \cite{5453745}. The scheduling policy can be redefined as $S(t) \in \mathcal{S} = \{0,1,...,L\}$, where $S(t) = l$ means that the nodes in cluster $l$ are scheduled to transmit at time slot $t$.
 \vspace{-1mm}

\subsection{Markov Decision Processes Formulation}

We formulate the problem as a Markov Decision Process (MDP) that is composed of the tuple $\langle s,a,r,p \rangle$, where $s$ is the state, $a$ presents the action, $r$ denotes the reward function, and $p$ describes the state transition probability. Hence, at time instant $t$, the agent (UAV) observes the current state $s(t)$ from the environment and tries to follow the optimal policy by selecting the best action $a(t)$, which maximizes the reward $r(t)$ and transiting to the next state $s(t+1)$ with a probability $p(s(t),s(t+1))$. For convenience, we propose an episodic MDP, where an episode starts with each UAV at one of the charging depots and ends when at least one UAV needs to recharge its battery at the nearest charging depot.

\subsubsection{State space}
The state space of the system at time slot $t$ is defined as $s(t) = (\boldsymbol{c}(t),\boldsymbol{A}(t), \boldsymbol{\beta}(t))$ where $\boldsymbol{c}(t)$  is a vector containing the position  of each UAV $c_u(t)\in\mathcal{C}$ at time slot $t$. $\boldsymbol{A}(t) = (A_1(t), A_2(t),...,A_L(t))$ contains the average AoI of the IoT devices in each cluster, where $A_l(t)\in\mathcal{I} = [1,2,...,A_{max}]$. $\boldsymbol{\beta}(t) = (\beta_1(t), \beta_2(t),...,\beta_U(t))$ with $\beta_U(t)\in \mathcal{B}$, is a vector that contains the difference between the battery status of each UAV  and both the required energy to arrive to the nearest charging depot $d \in \mathcal{D}$ and the energy consumed by packet relays considering the worst case when the UAVs relay packets in every time slot $t$. Finally, the state space of the system is  given by $\Sigma= \mathcal{C}^{U}\times\mathcal{I}^K\times\mathcal{B}^U$.

\subsubsection{Action space}
The action space at time slot $t$ is defined as $a(t) = (F_u(t),\mathcal{S}_u(t)$, where $f_u(t)$ is the movement of UAV $u$ and $\mathcal{S}_u(t)$ is the scheduling policy of UAV $u$. Each UAV $u$ selects a cluster $l$ to serve all the devices within this particular cluster. The action space is given by $\mathcal{A} = \mathcal{F}^U\times\mathcal{S}^U$. 

\subsubsection{Transition probability}
The transition between states relies on the 3 components of the state space. The AoI is updated according to \eqref{AOI_CALC}, the $\boldsymbol{\beta}$ is updated according to the energy calculations discussed in \ref{ENERGY_CALC}. The position of each UAV $c_u$ is updated according to the selected action $f_u(t)$, where \vspace{-2mm}
\begin{equation} \label{eqn:directions}
	c_u(t+1)=
	\begin{cases}
		c_u(t)+(0,d_g), & \quad f_u(t)=\text{North}, \\
		c_u(t)-(0,d_g), & \quad f_u(t)=\text{South}, \\
		c_u(t)+(d_g,0), & \quad f_u(t)=\text{East}, \\
		c_u(t)-(d_g,0), & \quad f_u(t)=\text{West}, \\
		c_u(t), & \quad \text{Hovering}. \\
	\end{cases}
\end{equation}

\subsubsection{Reward function}
The reward system is defined to minimize the weighted sum of the age of information as well as the average transmit power for all IoT devices. We define the immediate reward $r_u$ for the $u$ UAV at time instant $t$ as \vspace{-1mm}
\begin{equation}
r_u(t) = - \sum_{k = 1}^{K}\delta_k A_k(t) \: - \lambda \: \frac{1}{K} \sum_{k = 1}^{K} P_k, \label{eqn:reward}
\end{equation}
which is the DRL version of the objective function in \eqref{P1:a}.
\vspace{-1mm}
\subsection{DQN solution}
The state-action value function (Q-function) $Q_\pi(s,a)$ describes how good an action $a$ is at state $s$ while following the policy $\pi$~\cite{ADB+17_DRLsurvey}. It can be updated each time instant as follows
\begin{align}
& Q\left(s\left(t\right),a\left(t\right)\right) = \:  Q\left(s\left(t\right),a\left(t\right)\right) + \nonumber\\
&\alpha \:  \left(r\left(t\right) +  \gamma \: \max_a Q\left(s\left(t+1\right),a\right)
-Q\left(s\left(t\right),a\left(t\right)\right)\right),
\end{align} 
where $\alpha$ is the learning rate, $r(t)$ is the immediate reward, $\gamma \: Q\left(s\left(t+1\right),a\left(t+1\right)\right)$ is the discounted state-action value at time instant $t+1$, and $\gamma$ is the discount factor. 

\begin{figure}[t!]
	\centering
	\includegraphics[trim={1cm 2cm 1cm 2cm},clip,width=0.38\textwidth]{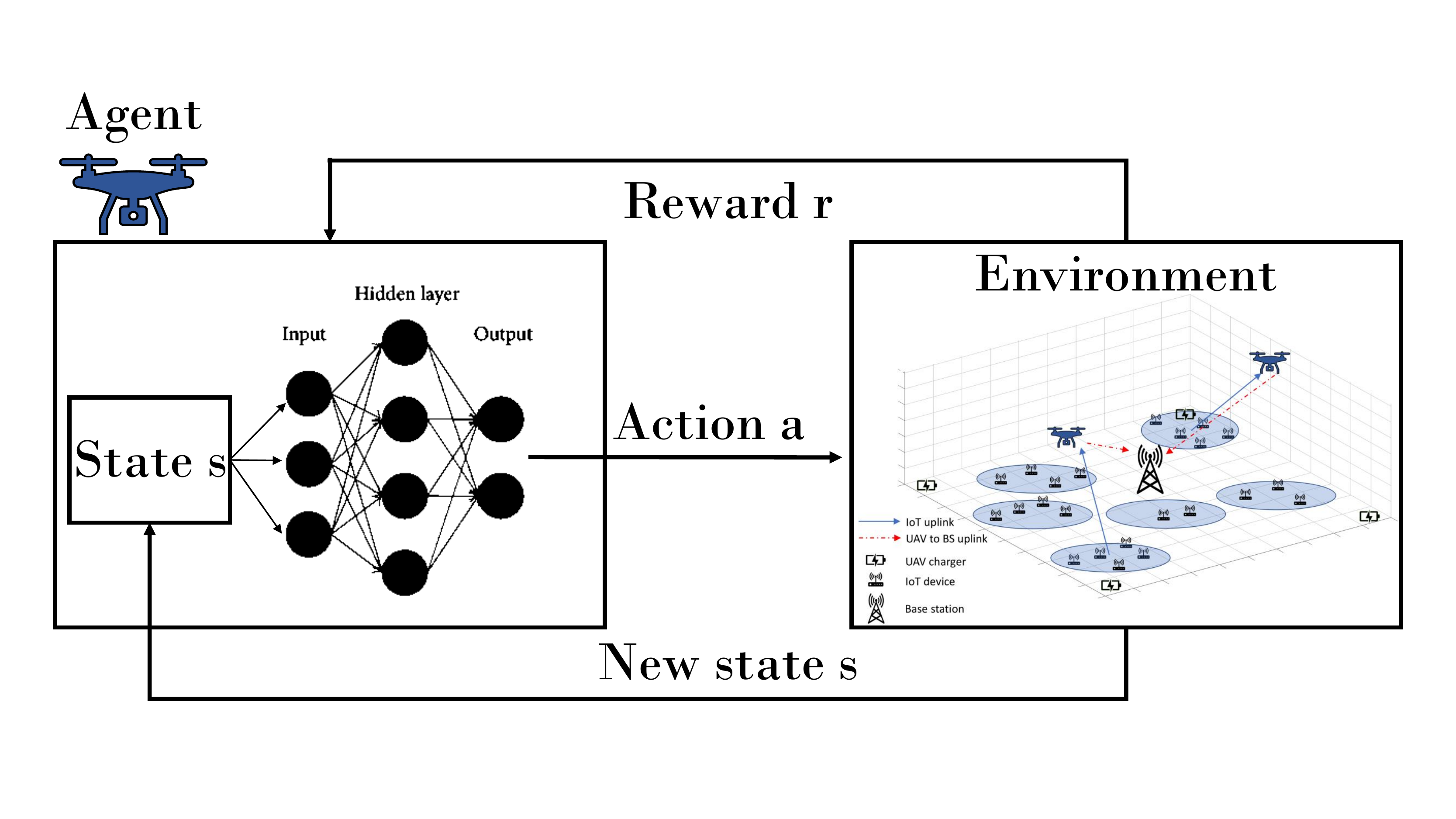}
	\caption{The DQN architecture.}
	\label{DQN_ARCH} \vspace{-2mm}
\end{figure}

The DQNs consist of two neural networks, where the first network (current network) works as a Q-function estimator, whereas the other (target network) works as a target Q-function network \cite{DQNs}. This approach solves the problem of large dimensionality in complex models. Moreover, the model defines the exploration rate $\epsilon$, which decays with time. To break the correlation between samples and utilize past samples, the DQN introduces experience replay, where it stores the past experiences $\langle s(t),a(t),r(t),s(t+1)\rangle$ in a buffer and samples a small batch randomly for training. Algorithm 1 summarizes the proposed DRL framework and Fig.~\ref{DQN_ARCH} illustrates the DQN architecture and interaction with the environment.
\vspace{-1mm}

\begin{algorithm}[!t]
\SetAlgoLined
Define parameters from table  \ref{tab:uav}.

Calculate $n_l$ using \eqref{eqn:rate_1}.

The number of clusters $L=\frac{K}{n_l}$.

Apply k-means to perform clustering.

Initialize the replay buffer and $t=1$.

Define $\epsilon$, $\gamma$, $\alpha$, $O$, and the number of episodes $E$.

Choose a value for $\lambda$ in \eqref{eqn:reward}.

\For{e = 1,...,$E$}{
    \While{No recharging needed (i.e. $\beta_1(t) > 0$),}{
        Explore a random action $a$ with probability $\epsilon$ or select optimal action $a = \max_a Q(s(t),a)$ with probability $1-\epsilon$.
        
        Save $\langle s(t),a(t),r(t),p(t) \rangle$ in the replay buffer.
        
        Sample a mini-batch from the buffer.
        
        Update the current network.
        
        Update the target network every $O$ instants.
        
        $t = t+1$.
    }
}

\caption{The proposed DRL algorithm}
\label{alg1}  \vspace{-1mm}
\end{algorithm}
 \vspace{-1mm}
\section{Numerical Results}\label{results}



In this section, we discuss the simulation results of the proposed DRL algorithm and compare them to various baseline models such as the GA, NN, RW. The GA tends to minimize the age only by scheduling and moving towards clusters with the highest age. This almost corresponds to the case when $\lambda=0$, and the UAV applies time division multiple access (TDMA) to distribute resources fairly. The NN always schedules the nearest cluster in order to minimize the transmit power. We consider a grid world of $1100$ m $\times$ $1100$ m, which is divided into $11 \times 11$ grids. The simulation parameters are defined in Table \ref{tab:uav}. 

We build a DQN of five hidden layers (64,128,256,128,128 neurons) with $\alpha = 0.0001$, Adam optimizer, replay buffer of size $100000$, $\gamma = 0.99$, and $100000$ trained episodes using Pytorch framework on NIVIDIA Tesla V100 GPU. The proposed DQN model has spatial complexity illustrated in terms of the number of parameters (weights and biases) of $344,290$ parameters, which need around $30 MB$ of memory. In terms of the computational complexity, the model performs $170,816$ multiplications and additions. The time complexity to execute one episode using the proposed algorithm is $0.0918 \: s$ compared to the $0.0665 \: s$ of the RW. Throughout this section, the term "ergodic" refers to time and statistical average.
\begin{table}[t!]
\centering
\caption{UAV model parameters}
\label{tab:uav}
\begin{tabular}{cc|cc|cc}
\toprule
\textbf{Parameter}                                    & \textbf{Value} & \textbf{Parameter}                                    & \textbf{Value} & \textbf{Parameter}                                    & \textbf{Value}\\ \midrule
$E_{max,u}$ & 10000 & $e_{max,u}$ & 200 & $A_{max}$ & 30 \\

$g_0$ & 30 dB & $h_u$ & 100 m & $d_g$ & 100 m \\

$B$ & 1 MHz & $M$ & 5 Mb  & $\sigma^2$ & -100 dBm \\

$\mathcal{C}$ & 4 & $\upsilon_t$ & 25 m/s & $s_{tip}$ & 120 m/s \\

$\rho$ & 1.225 kg/m$^3$ & $P_0$ & 99.66 W & $P_1$ & 120.16 W \\

$d_0$ & 0.48 & $\mu_0$ & 0.0001  & $Z$ & 0.5 s$^2$  \\

$s_0$ & 0.002 m/s & $h_{BS}$ & 15 m\\
\bottomrule
\end{tabular} 
\end{table}

\begin{figure*}[t!]
    \centering
    \subfloat[NN\label{Fig_2a}]{\includegraphics[width=0.31\textwidth,trim={4cm 0 5cm 0},clip]{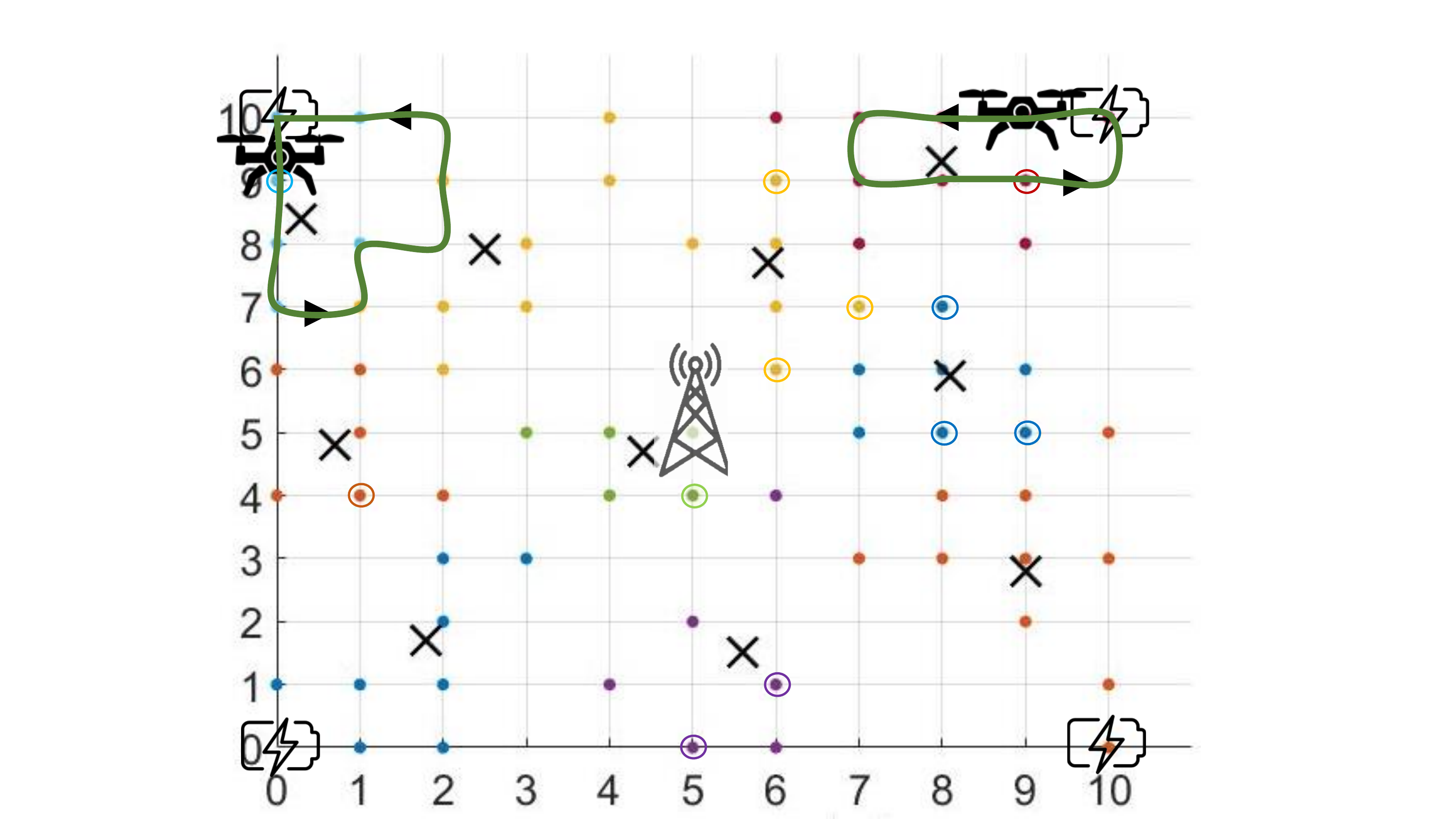}}
    \hskip -2.9ex
    \subfloat[GA\label{Fig_2b}]{\includegraphics[width=0.31\textwidth,trim={4cm 0 5cm 0},clip]{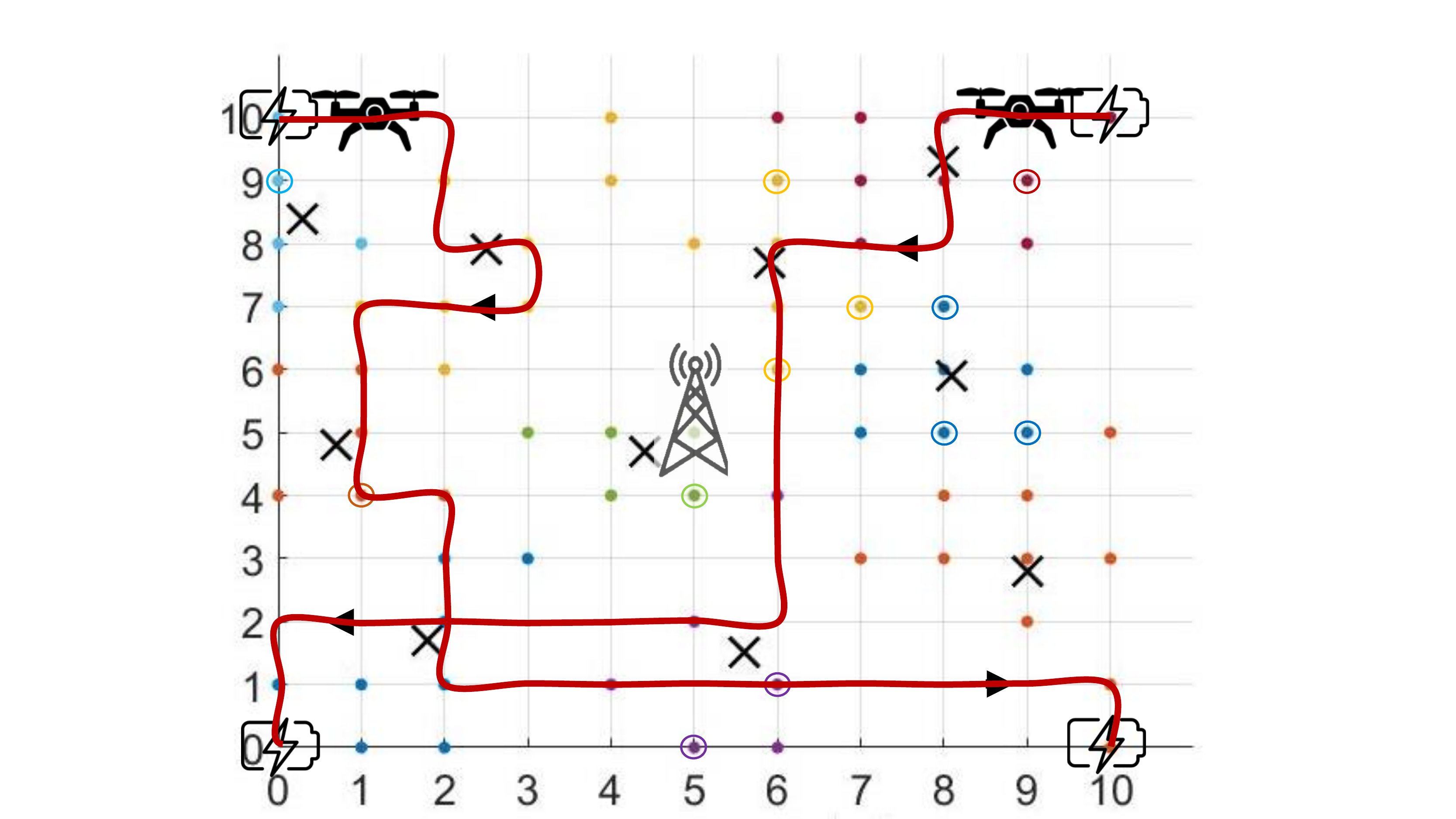}}
    \hskip -2.9ex
    \subfloat[DRL\label{Fig_2c}]{\includegraphics[width=0.31\textwidth,trim={4cm 0 5cm 0},clip]{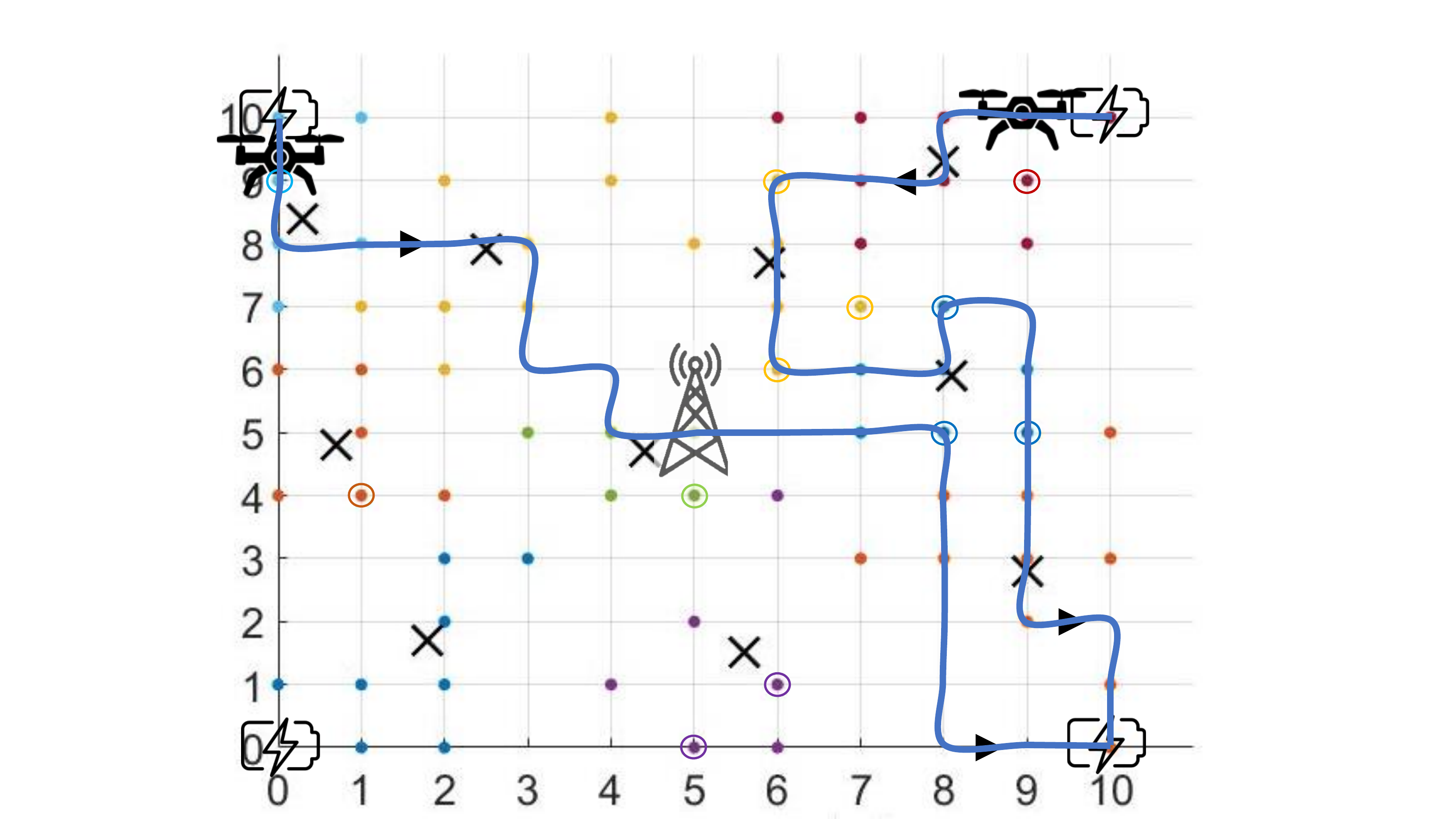}}
    \hskip -2.9ex
    \caption{Trajectories for $K = 100$ and $U = 2$ at a trained episode. For the DRL scheme, $\lambda = 25$. The colored points represent IoT devices, where different clusters are indicated by different colors. Crosses represent the cluster centroids. Circled points indicate the presence of multiple devices at those coordinates.}
    \label{Fig2} \vspace{-2mm}
\end{figure*}

Figure \ref{Fig2} presents an example trajectory path of two UAVs for a trained episode. We can notice that with the NN in Fig.~\ref{Fig_2a}, the UAVs move randomly and schedule the nearest devices. In Fig.\ref{Fig_2b}, the GA chooses the devices with the highest age careless of the large path losses. Fig.~\ref{Fig_2c} shows the trained DRL scheme. Since more devices are located in the right upper section of the map, both UAVs tend to fly over the cluster centroids close to this region, which indicates the learning behaviour. Moreover, it is worth concluding that a free flight passing above these centroids could be a low-complexity sub-optimal trajectory. 



\begin{figure*}[t!]
    \centering
    \subfloat[Accumulative reward\label{Fig_3a}]{\includegraphics[width=0.3\textwidth]{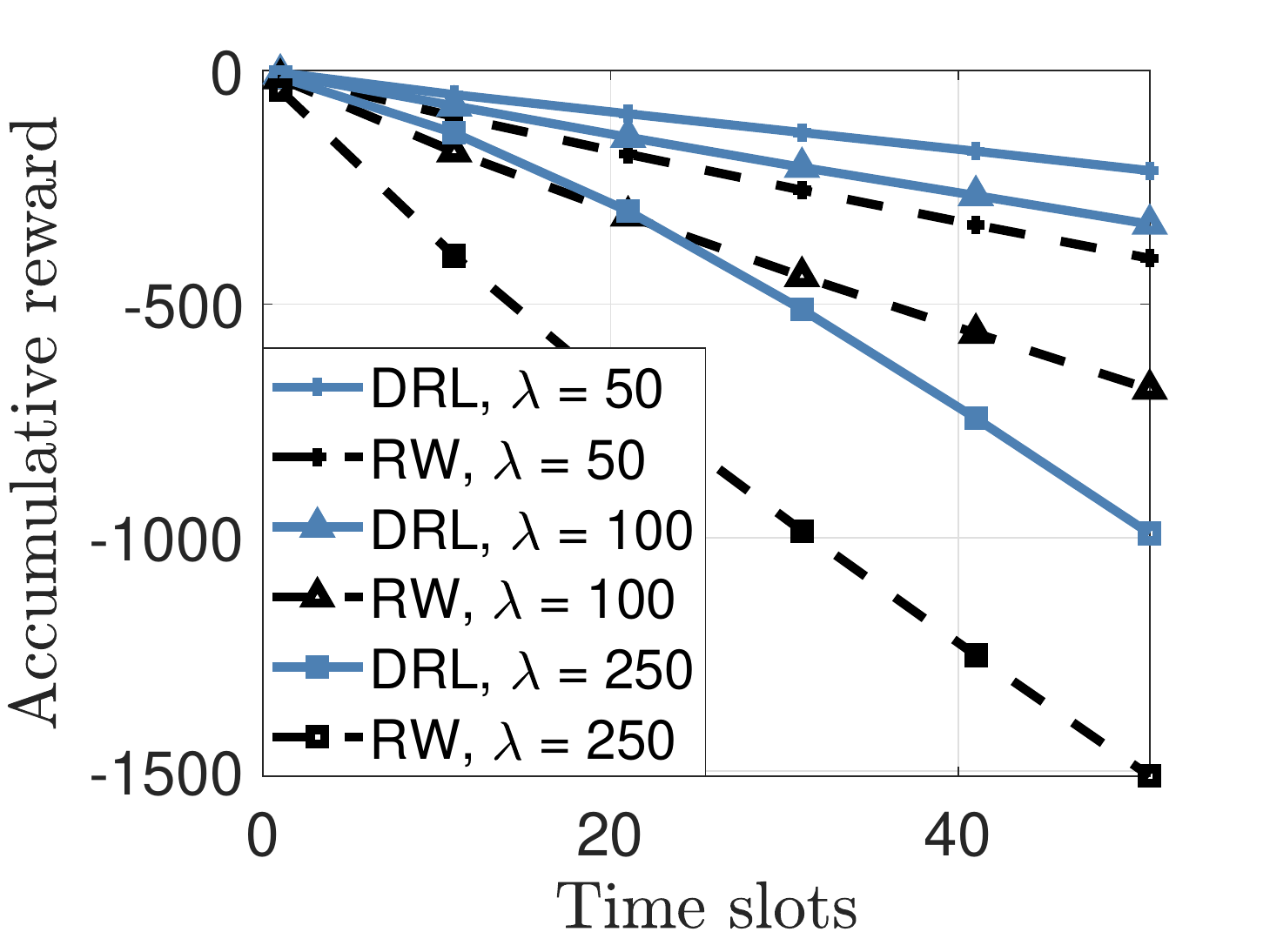}}
    \hskip -1.9ex
    \subfloat[Ergodic age\label{Fig_3b}]{\includegraphics[width=0.3\textwidth,trim={0 0 0 0},clip]{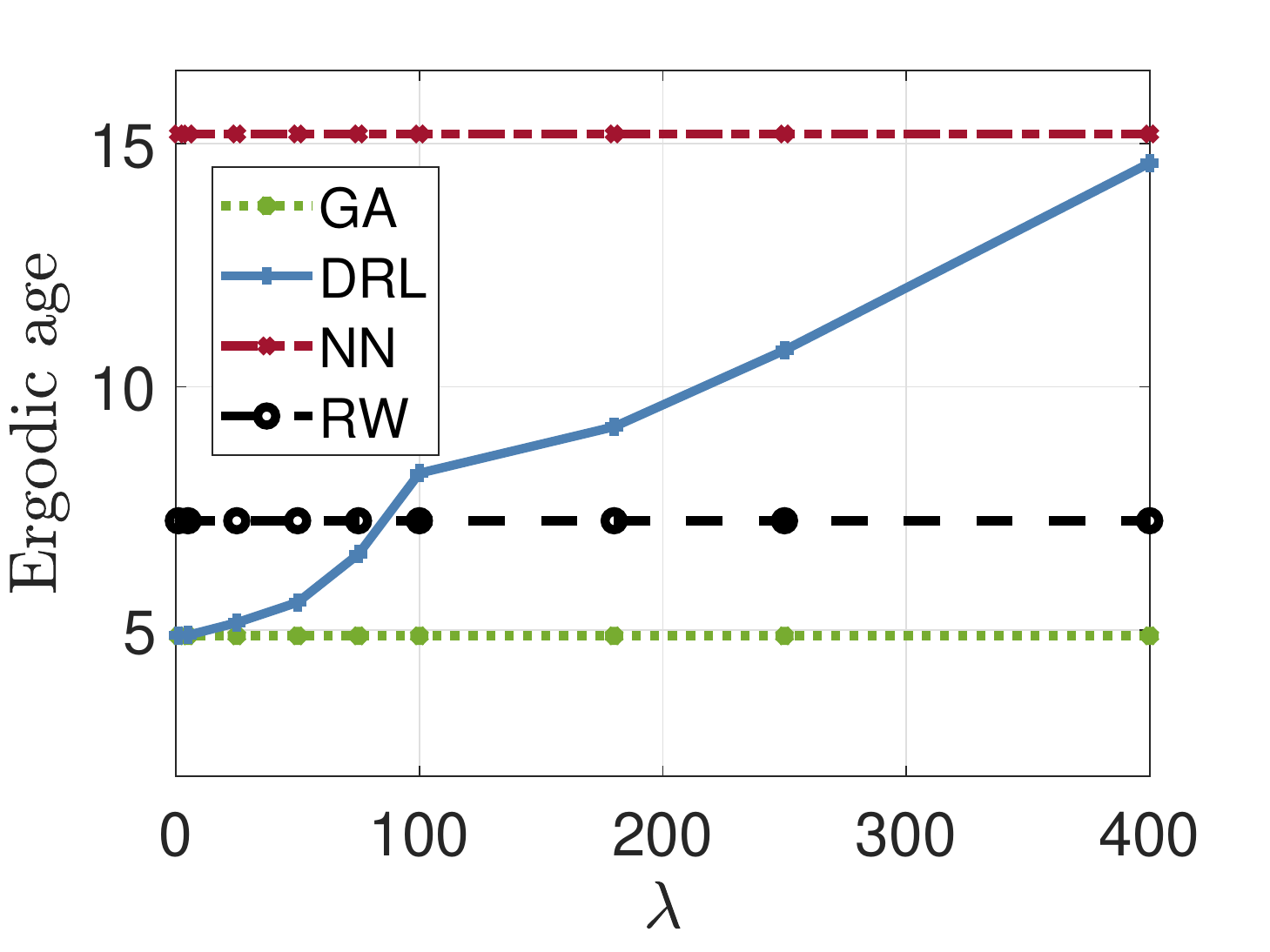}}
    \hskip -1.9ex
    \subfloat[Ergodic power\label{Fig_3c}]{\includegraphics[width=0.3\textwidth,trim={0 0 0 0},clip]{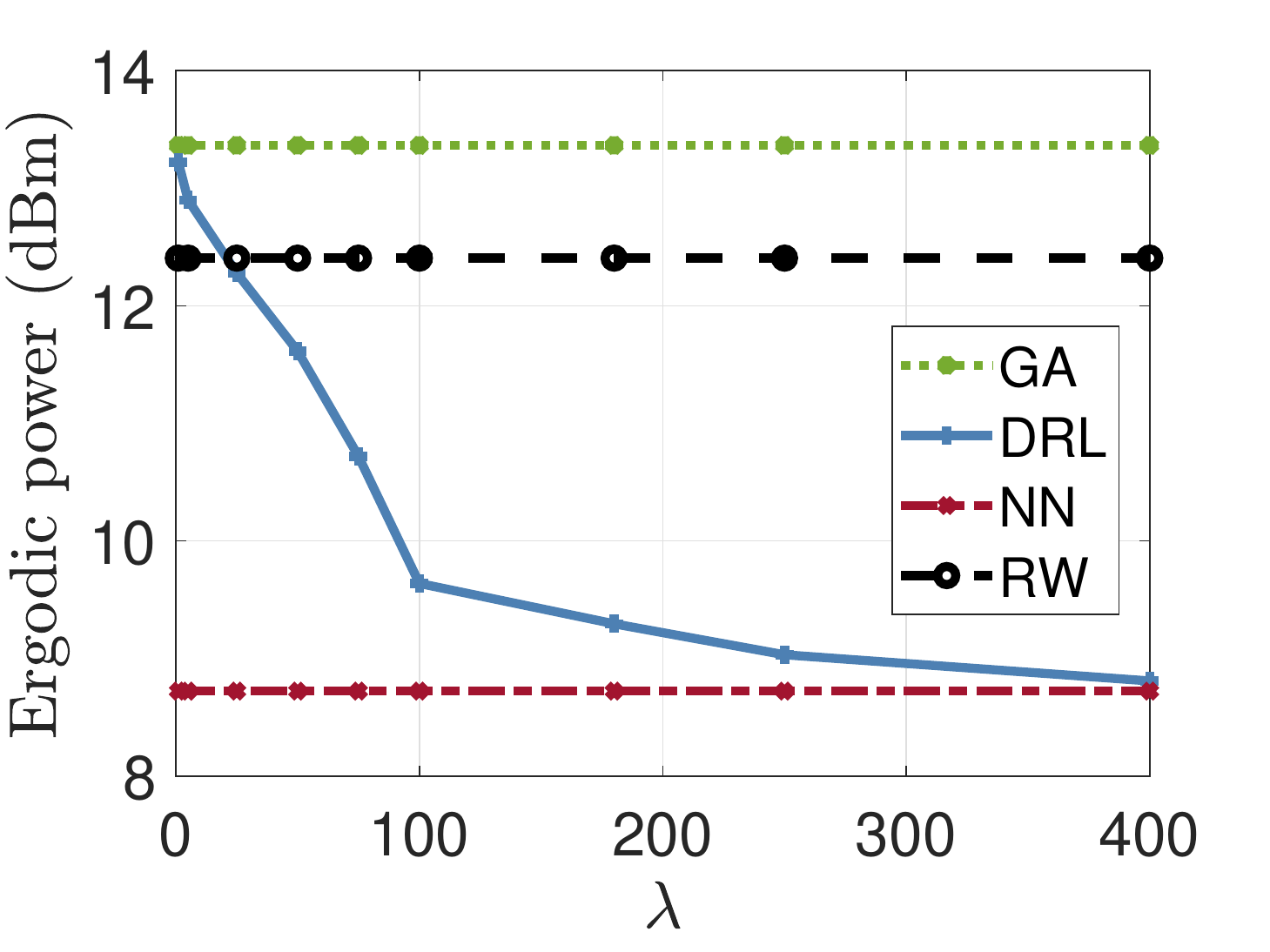}}
    \hskip -1.9ex
    \caption{Accumulative reward, ergodic age, and ergodic power for the GA, DRL, NN, and RW schemes at $K=100$, and $U = 2$.}
    \label{Fig_3} \vspace{-2mm}
\end{figure*}

\begin{figure}[t!]
	\centering
\includegraphics[width=0.38\textwidth]{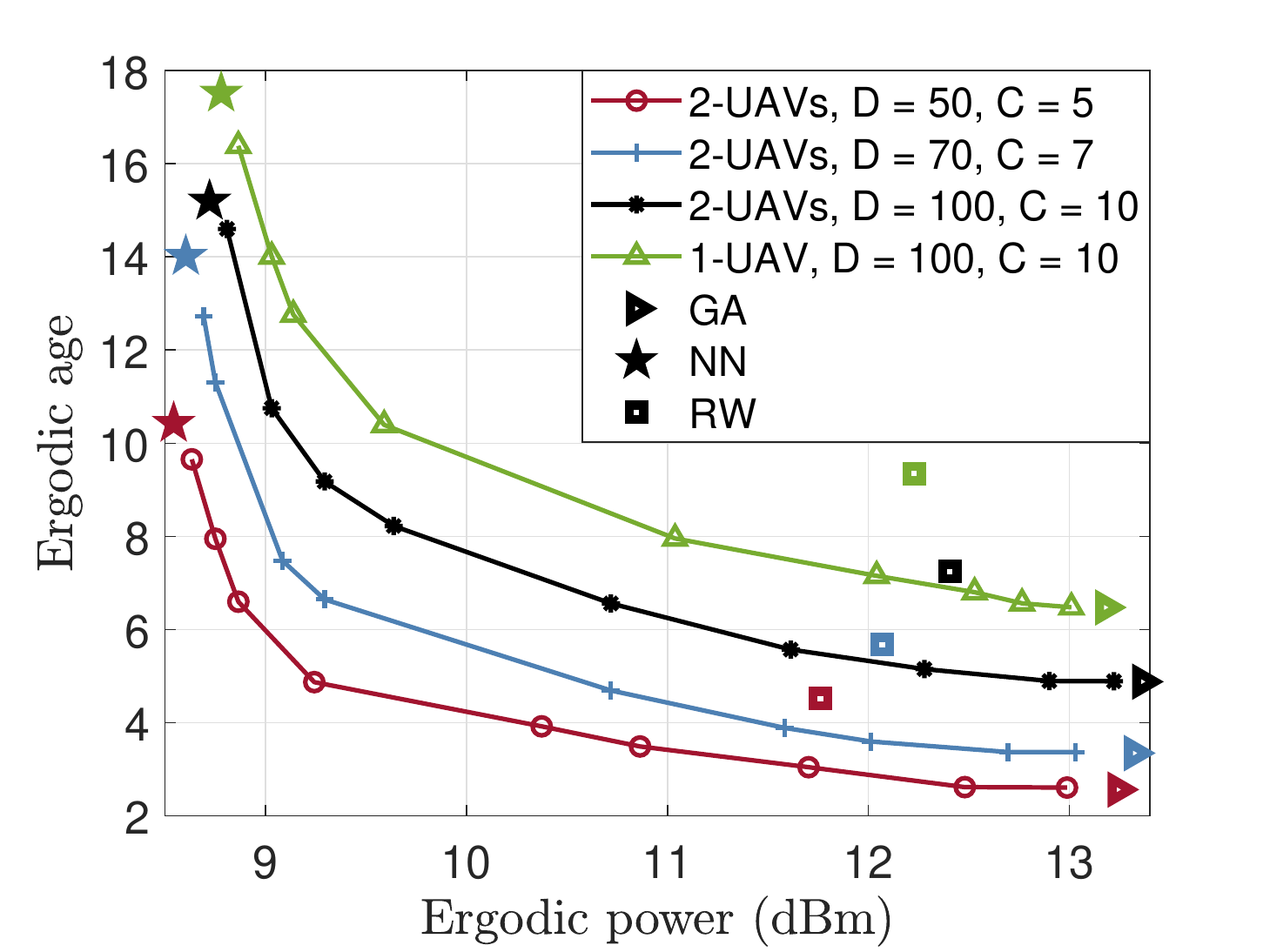} \vspace{-0.5mm}
	\caption{Achievable region of ergodic age and ergodic power for the DRL, GA, NN, and RW schemes, adjusting the values of $\lambda$, $U$, $D$, and $C$.}
	\label{Fig4} \vspace{-3mm}
\end{figure}

Figure \ref{Fig_3a} depicts the accumulative reward for the DRL and RW schemes for different values of $\lambda$. It is not a surprise that higher $\lambda$ values reflect lower accumulative rewards due to the nature of the reward function in \eqref{eqn:reward}. However, we can see that the DRL scheme offers a significant improvement in the reward compared to the RW for all $\lambda$ values. Looking at figures \ref{Fig_3b} and \ref{Fig_3c}, it was also expected that neither the age nor the power consumption are affected by $\lambda$ for all schemes expect for the DRL. This present an aspect of adaptability for the DRL scheme, where one can choose to prioritize the age or the power consumption, and vice versa using the same algorithm. Thus, it can achieve promising results on the age as the GA scheme, or lower power consumption as the NN scheme. This exchange can be observed in Fig. \ref{Fig4}, where we observe the achievable regions of age and power for the DRL scheme for different values of $\lambda$ values. We can see the DRL scheme as lines, since it benefits from the variation of $\lambda$, where the other schemes are just static points. Another important insight is that increasing the number of UAVs as well as decreasing the number of IoT devices improve the values of both age and transmit power in the achievable region.

\vspace{-1mm}
\section{Conclusions}\label{conclusions} \vspace{-1mm}
In this paper, we considered a relatively large IoT network, where multiple UAVs serve as mobile relay nodes with the objective of minimizing the age of information and the energy consumption. The problem was formulated as an optimization problem to plan the trajectory of the UAVs from one charging depot to another such that the ergodic age and energy consumption of the network is minimized. We addressed the problem by proposing a DRL-based solution, where the BS clusters the IoT devices according to their positions and UAV flight time between grids to improve the performance. Our proposed approach outperforms other state-of-the-art solutions such as GA, NN and RW. In particular, the proposed DRL-based solution provides the best age-energy trade-off in a wide range of scenarios involving different numbers of UAVs and IoT nodes. Another contribution of this work is the simplicity of the proposed solutions, which addresses the problem of high dimensionality in the action space, thus enabling its application in a massive IoT deployment scenario with the number of IoT devices in the hundreds as a future extension. %
\vspace{-2mm}



%
%

%
\bibliographystyle{IEEEtran}
\bibliography{di}

\begin{thebibliography}{10}
\providecommand{\url}[1]{#1}
\csname url@samestyle\endcsname
\providecommand{\newblock}{\relax}
\providecommand{\bibinfo}[2]{#2}
\providecommand{\BIBentrySTDinterwordspacing}{\spaceskip=0pt\relax}
\providecommand{\BIBentryALTinterwordstretchfactor}{4}
\providecommand{\BIBentryALTinterwordspacing}{\spaceskip=\fontdimen2\font plus
\BIBentryALTinterwordstretchfactor\fontdimen3\font minus
  \fontdimen4\font\relax}
\providecommand{\BIBforeignlanguage}[2]{{%
\expandafter\ifx\csname l@#1\endcsname\relax
\typeout{** WARNING: IEEEtran.bst: No hyphenation pattern has been}%
\typeout{** loaded for the language `#1'. Using the pattern for}%
\typeout{** the default language instead.}%
\else
\language=\csname l@#1\endcsname
\fi
#2}}
\providecommand{\BIBdecl}{\relax}
\BIBdecl

\bibitem{kaul2011minimizing}
S.~Kaul, M.~Gruteser, V.~Rai, and J.~Kenney, ``Minimizing age of information in
  vehicular networks,'' in \emph{8th Annual IEEE Communications Society
  Conference on Sensor, Mesh and Ad Hoc Communications and Networks}.\hskip 1em
  plus 0.5em minus 0.4em\relax IEEE, 2011, pp. 350--358.

\bibitem{tang2020minimizing}
H.~Tang, J.~Wang, L.~Song, and J.~Song, ``Minimizing age of information with
  power constraints: Multi-user opportunistic scheduling in multi-state
  time-varying channels,'' \emph{IEEE Journal on Selected Areas in
  Communications}, vol.~38, no.~5, pp. 854--868, 2020.

\bibitem{mozaffari2019tutorial}
M.~Mozaffari \emph{et~al.}, ``A tutorial on {UAVs} for wireless networks:
  Applications, challenges, and open problems,'' \emph{IEEE communications
  surveys \& tutorials}, vol.~21, no.~3, pp. 2334--2360, 2019.

\bibitem{DQNs}
V.~Mnih \emph{et~al.}, ``Human-level control through deep reinforcement
  learning,'' \emph{Nature}, vol. 518, pp. 529--33, 02 2015.

\bibitem{abd2019deep}
M.~A. Abd-Elmagid, A.~Ferdowsi, H.~S. Dhillon, and W.~Saad, ``Deep
  reinforcement learning for minimizing age-of-information in {UAV}-assisted
  networks,'' in \emph{2019 IEEE GLOBECOM}, 2019, pp. 1--6.

\bibitem{9195789}
M.~Samir \emph{et~al.}, ``Age of information aware trajectory planning of uavs
  in intelligent transportation systems: A deep learning approach,'' \emph{IEEE
  Transactions on Vehicular Technology}, vol.~69, no.~11, pp. 12\,382--12\,395,
  2020.

\bibitem{9750860}
O.~S. Oubbati \emph{et~al.}, ``{Synchronizing UAV Teams for Timely Data
  Collection and Energy Transfer by Deep Reinforcement Learning},'' \emph{IEEE
  Transactions on Vehicular Technology}, pp. 1--1, 2022.

\bibitem{samir2020online}
M.~Samir, C.~Assi, S.~Sharafeddine, and A.~Ghrayeb, ``Online altitude control
  and scheduling policy for minimizing {AoI in UAV-assisted IoT} wireless
  networks,'' \emph{IEEE Transactions on Mobile Computing}, 2020.

\bibitem{ferdowsi2021neural}
A.~Ferdowsi~et al, ``Neural combinatorial deep reinforcement learning for
  age-optimal joint trajectory and scheduling design in {UAV}-assisted
  networks,'' \emph{IEEE Journal on Selected Areas in Communications}, vol.~39,
  no.~5, pp. 1250--1265, 2021.

\bibitem{deep_us}
M.~A. Abd-Elmagid, A.~Ferdowsi, H.~S. Dhillon, and W.~Saad, ``Deep
  reinforcement learning for minimizing age-of-information in {UAV}-assisted
  networks,'' in \emph{2019 IEEE GLOBECOM}, 2019, pp. 1--6.

\bibitem{deep_china}
M.~Yi \emph{et~al.}, ``Deep reinforcement learning for fresh data collection in
  {UAV}-assisted {IoT} networks,'' in \emph{IEEE INFOCOM Workshops 2020}, 2020,
  pp. 716--721.

\bibitem{zeng2019energy}
Y.~Zeng, J.~Xu, and R.~Zhang, ``Energy minimization for wireless communication
  with rotary-wing {UAV},'' \emph{IEEE Transactions on Wireless
  Communications}, vol.~18, no.~4, pp. 2329--2345, 2019.

\bibitem{5453745}
S.~Na, L.~Xumin, and G.~Yong, ``Research on k-means clustering algorithm: An
  improved k-means clustering algorithm,'' in \emph{2010 Third International
  Symposium on Intelligent Information Technology and Security Informatics},
  2010, pp. 63--67.

\bibitem{ADB+17_DRLsurvey}
K.~Arulkumaran, M.~P. Deisenroth, M.~Brundage, and A.~A. Bharath, ``Deep
  reinforcement learning: A brief survey,'' \emph{IEEE Signal Processing
  Magazine}, vol.~34, no.~6, pp. 26--38, Nov. 2017.

\end{thebibliography}
\end{document}